\documentclass[journal]{IEEEtran}

\usepackage{cite}
\usepackage{amsmath}
\usepackage{amsfonts}
\usepackage{graphicx}

\ifCLASSINFOpdf
\else
\fi

\begin{document}

\title{Analysis of Geometric Eigenmode Splitting in Planar Microwave Resonators}

\author{Hagen~Papenburg,~
        Jason~Corn,~
        Romain~Speciel,~
        and~Adham~Naji
\thanks{H.~Papenburg and J.~Corn are with the Department
of Electrical and Computer Engineering, Santa Clara University, Santa Clara,
CA, 95053 USA.}
\thanks{R.~Speciel is with the Mathematics Department, Stanford University, Stanford, CA, 94305 USA.}
\thanks{A.~Naji is with the Department
of Electrical and Computer Engineering, the Department of Physics, and the Department of Applied Mathematics, Santa Clara University, Santa Clara,
CA, 95053 USA (Corresponding author: Adham Naji; anaji@scu.edu).}
}


\maketitle

\begin{abstract}
A traditional challenge in planar microwave resonator design is to find geometric mechanisms that can control individual resonant frequencies --- or tune coupling between specific eigenmodes --- without affecting other system parameters, such as the center frequency.
Achieving this capability can be transformative for the design of microwave sensors, filters, antennas and resonant networks.
Starting from first principles, we outline the development of an analytical method that guides practical design by suggesting specific perturbation shapes capable of achieving resonant mode-splitting while keeping the center frequency nearly stationary.
This approach opens the door to novel designs that were previously difficult to discover or optimize.
Independent numerical simulations and measurements of stripline resonators validate the effectiveness of this approach. The demonstrated tuning effect shows that the first-order estimate suggested by the theory persists in practice over a relatively wide splitting bandwidth (exceeding 54\% of center frequency), increasing the utility of the proposed method.
\end{abstract}

%
\IEEEpeerreviewmaketitle

\section{Introduction}

Electromagnetic resonators represent the basic building blocks for many microwave and radio frequency (RF) sensors, filters, antennas, charged-particle accelerators, and guided-wave networks \cite{Collin, Hong, Cameron,Balanis, Slater}. 
Operating in the microwave regime---where characteristic resonator dimensions are comparable to the guided wavelength---these systems exhibit a rich variety of wave phenomena that are obscured or absent in the asymptotic limits of lower or higher frequencies (e.g., lumped-element circuit theory or geometrical optics). The analysis of these systems, however, has traditionally suffered from the complex relationship between the geometric shape of the  resonator hardware and its corresponding frequency response. 
A typical example occurs when attempting to detune a specific eigenmode or control inter-mode coupling in a resonant network; perturbations can inadvertently shift other resonant frequencies across the spectrum and drift the center frequency of the ensemble \cite{Hong, Cameron, Collin, Naji_geo, Naji_stability, Hong_triangle, Gorur,Lin}. This complexity stems from the spectral geometry of the governing Helmholtz boundary-value problems, which are often challenging to treat analytically. Many practical designs therefore rely on numerical simulations or traditional trial-and-error methods. 
Establishing a systematic theoretical framework to leverage the relationship between the eigen-spectrum and the geometry will provide a powerful toolkit that informs practical design.

\begin{figure}[t]
\centering
\includegraphics[width=75mm]{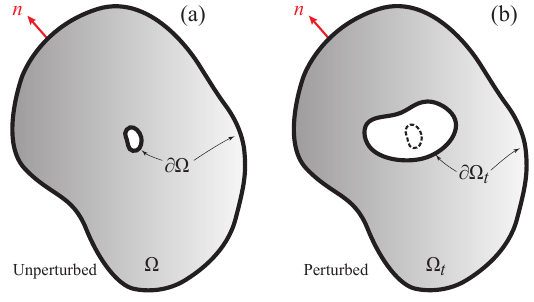}
\caption{An illustration of the cross section $\Omega$ of an arbitrary planar resonator before perturbation (a) and $\Omega_t$ after perturbation (b).}
\label{fig:resonator}
\vspace{-\baselineskip}
\end{figure}

In this paper, we develop a framework for addressing this complexity in planar resonators, by rigorously analyzing how the geometry of a perturbation can be linked to the eigenfrequencies of modes of interest in the resonator. For general background on shape perturbation theory and on spectral geometry applied to electromagnetic modeling, we refer to \cite{Henry, Delfour, Lamberti}. The level of geometric spectral control offered in the present paper, even to first order, can open the door to novel and previously unexplored parts of the solution space of possible designs. Specifically, in the following sections we demonstrate how we can analytically find a family of perturbation shapes that manipulate the splitting of degenerate resonant frequencies in a planar resonator, without changing the center frequency at first order. This is not only of theoretical interest, but also of great practical importance in sensor, filter, and antenna applications. For example, in coupled-resonator networks, one often needs to tune mode coupling values while keeping the shared reference frequency fixed to avoid internal network mismatches that degrade the frequency response\cite{Naji_stability,Cameron, Hong}. In reconfigurable dual-band systems, one may want to sweep the bands while keeping the center frequency unchanged around a prescribed carrier \cite{Cameron, Hong}. The underlying analysis method can be further extended to achieve different spectral design goals in other scenarios. \\

Starting from first principles, the theoretical formulation of the eigen-analysis is summarized in Section~\ref{sec: theory}. To verify the theory, we conduct measurements and utilize independent Finite Element Method (FEM) simulations for a stripline resonator design in Section~\ref{sec: examples}. Good agreement is demonstrated.


\section{Theoretical Formulation}\label{sec: theory}

We model a microstrip resonator as a cavity $\Omega \times [0,h]$, where $\Omega\subseteq \mathbb{R}^2$ is an arbitrary domain in the $xy$ plane and $h$ is a small height along the $z$ axis (see Figure~\ref{fig:resonator}). The cavity has perfect electric conductor (PEC) boundary conditions ($\hat{n} \times \mathbf{E}=0$, $\hat{n} \cdot \mathbf{B} =0$) on the top and bottom surfaces and perfect magnetic conductor (PMC) boundary conditions ($\hat{n} \times \mathbf{B}=0$, $\hat{n} \cdot \mathbf{E} =0$) on the side walls. Here $\hat{n}$ denotes the outward pointing unit normal to the boundary, and $\mathbf{E},\mathbf{H}$ represent the phasors of the fields, where harmonic time-dependence $e^{j \omega t}$ is assumed.  
For transverse magnetic (TM) waves in a source-free, homogeneous, isotropic medium with the stated boundary conditions,
Maxwell's curl equations $\nabla\times\mathbf{E} + j \omega \tilde{\mu} \mathbf{H} =0$ and $\nabla\times\mathbf{H} - j \omega \varepsilon \mathbf{E} =0$ reduce to the scalar Helmholtz equation for the electric field component $E_z$ on the two-dimensional domain $\Omega$, subject to homogeneous Neumann boundary conditions:
\begin{equation}\label{Helmholtz}
    \Delta_{\perp}E_z + k_c^2 E_z = 0 \ \text{on} \ \Omega , \quad \partial_n E_z = 0 \ \text{on} \ \partial \Omega.
\end{equation}
Here $\Delta_{\perp}\equiv\nabla^{2}_{\perp}$ denotes the transverse Laplacian, $\partial_n$ denotes the normal derivative on the boundary $\partial\Omega$, and $k_c^2=k^2-k_z^2$ is the transverse cutoff wavenumber, with $k_z=p\pi/h$ being the longitudinal wavenumber ($p$ an integer) and $k=\omega/c$ being the wavenumber 
at angular frequency $\omega$ and speed of light $c=1/ \sqrt{\tilde{\mu} \varepsilon}$ in the medium with permeability $\tilde{\mu}$ and permittivity~$\varepsilon$. For microstrip or stripline cavity structures, where we take the height $h$ to be much smaller than the operating wavelength (thus $k<\pi/h$), we expect the first few eigenmodes of interest to be uniform along $z$, with $p=0$ and $k=k_c$. With $H_z=0$, one finds (see, for example, \cite{Collin}) that the TM H-field components can be obtained from $E_z$ as 
\begin{equation}\label{Hcomponents}H_x=\frac{j\omega \varepsilon}{k_c^2}\frac{\partial E_z}{\partial y}, \ \ H_y=\frac{-j\omega \varepsilon}{k_c^2}\frac{\partial E_z}{\partial x},\end{equation} 
which are out of phase by 90 degrees compared to $E_z$. 

Treating \eqref{Helmholtz} as an
eigenvalue problem for the negative Laplacian $-\Delta_{\perp}$, 
with eigenvalues
$\mu_j=k_c^2$ and eigenfunctions $u=E_z$, the spectrum on any bounded domain $\Omega \subset \mathbb{R}^2$ with smooth boundary is of the form  $0{=}\mu_0{<}\mu_1{\leq}\mu_2{\leq}\mu_3{\leq}\cdots$, and the lowest resonant frequency associated with an eigenvalue $\mu_j$ is $f_j=\frac{c\sqrt{\mu_j}}{2 \pi}$. We are
interested in analyzing the splitting behavior of degenerate resonant frequencies
under perturbations of the
shape $\Omega$.

For a smooth vector field $\mathbf{V}$ describing the change in shape, we represent the associated perturbation family over fictitious time $t$ by $\Omega_t=(\operatorname{Id}+t\mathbf{V})(\Omega)$, with $\operatorname{Id}$ denoting the identity transformation. Hadamard's variational formula \cite{Henry, Delfour} states that the derivative of a simple Neumann eigenvalue $\mu$, corresponding to unit norm eigenfunction $u$, is
\begin{equation}\label{simple}
    \frac{d}{dt}\Big|_{t=0}\  \mu (\Omega_t)
    =
    \int_{\partial\Omega}
    \left( |\partial_\tau u|^2-\mu u^2\right)\rho\,dS,
\end{equation}
where $\partial_\tau u$ denotes the tangential derivative of $u$ along the boundary, and $\rho = \mathbf{V}\cdot \hat{n}$ on $\partial \Omega$
is the normal displacement of the boundary.
An eigenvalue $\mu$ with multiplicity $m>1$ generically splits into several differentiable branches. Denoting by $\{u_1,\dots,u_m\}$
an $L^2(\Omega)$-orthonormal basis of the eigenspace $E_{\mu}$, the first derivatives of the eigenvalue branches are given by the eigenvalues of the symmetric $m\times m$ matrix
\begin{equation}\label{matrixentires}
    A_{ij}
    =
    \int_{\partial\Omega}
    \left(
        \partial_{\tau} u_i \partial_{\tau} u_j
        -
        \mu u_i u_j
    \right)\rho\,dS.
\end{equation}
The simple-eigenvalue formula \eqref{simple} is the special case $m=1$. \\

For an annulus resonator shape, the matrix entries
\eqref{matrixentires} can be calculated explicitly. We consider perturbations of the inner boundary, described in polar coordinates as
\begin{equation}\label{perturbations}
\Omega_t = \{(r, \theta) \ | \ a+t \cdot h(\theta) \leq r \leq b\}.
\end{equation}
The eigenvalues of the Helmholtz equation \eqref{Helmholtz} for the annulus $\Omega_0 = \{a \leq r \leq b\}$ can be labeled as $\mu_{m, n}=k_{m,n}^2$, $m \geq 0$, $n \geq 1,$ with associated eigenfunctions $u_{0,n}=R_{0,n}(r)$, and $u_{m,n}^c=R_{m,n}(r) \cos(m\theta)$, $u_{m,n}^s=R_{m,n}(r) \sin(m\theta)$ (with $m \geq 1$). The radial functions $R_{m,n}(r), n \geq 1$ are the solutions to the Sturm-Liouville problem consisting of $R''_{m,n}(r) +\frac{1}{r} R'_{m,n}(r) + \big(\mu_{m,n} - \frac{m^2}{r^2}\big)R_{m,n}(r)=0$ and the boundary conditions $R_{m,n}'(a)=R'_{m,n}(b)=0$. 
They can be expressed as a particular linear combination of rescaled Bessel functions of the first and second kind:
\begin{equation}\label{radialfunction}R_{m,n}(r)=Y_m'(k_{m,n}a)J_m(k_{m,n}r)-J_m'(k_{m,n}a)Y_m(k_{m,n}r),\end{equation}
where the admissible scaling factors $k_{m,n}$, with $n \geq 1$, are the positive zeros of the defining equation 
\begin{equation}\label{equationforks}
J_m'(k a)Y_m'( k b) - J_m'( k b)Y_m'( k a)=0,    
\end{equation}
 labeled in increasing order of $n$. For fixed $n$, the values of $k_{m,n}$ are also increasing in the first index $m \geq 1$. This can be shown by utilizing the variational characterization of eigenvalues via Rayleigh quotients \cite[section 6.2]{Collin}. ($m=0$ is the exception to this rule.)

\begin{figure}[b]
\centering
\includegraphics[width=88mm]{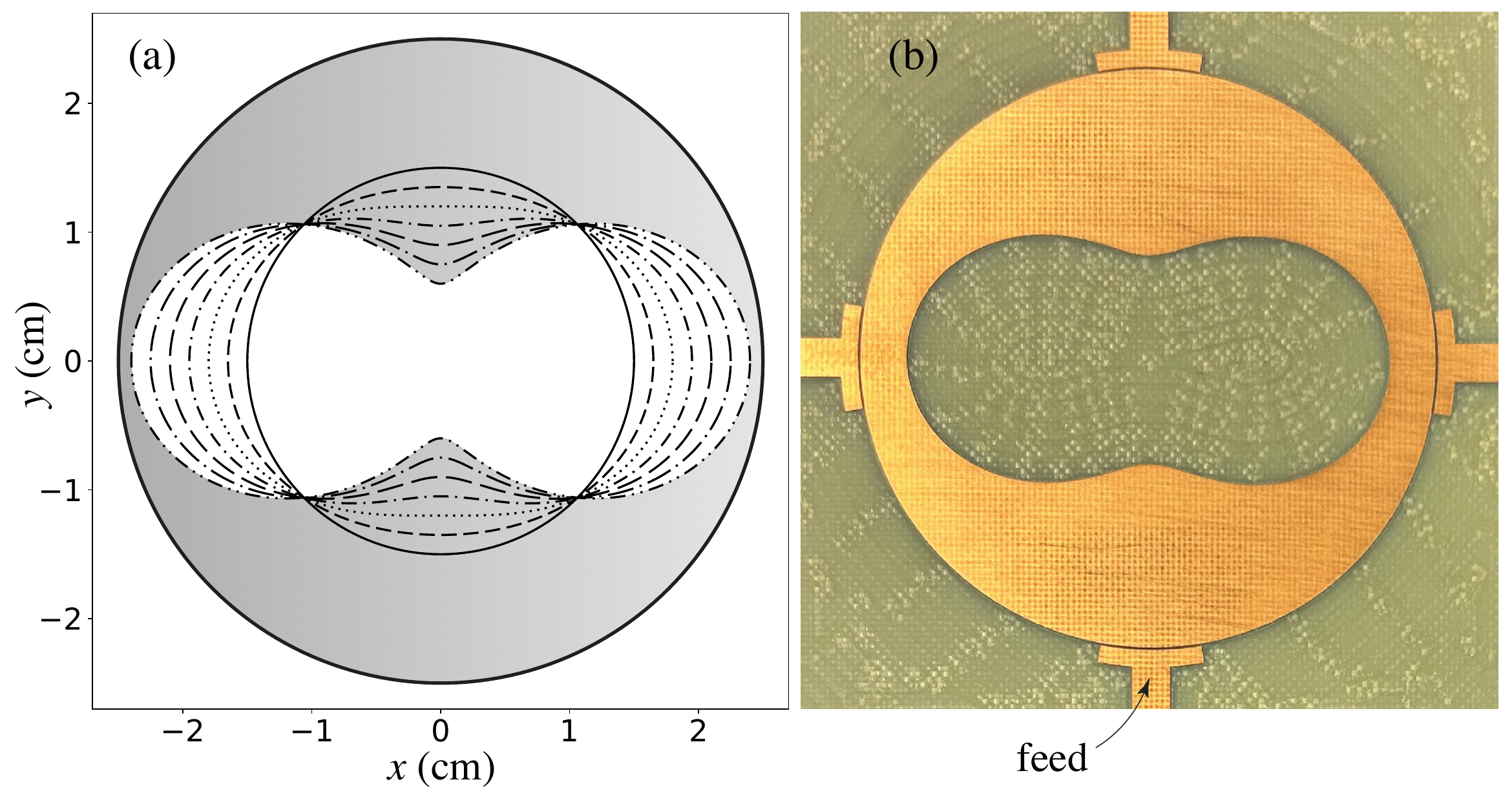}
\caption{(a) Inner-boundary perturbations, $r = a + t\cos(2\theta)$, with initial inner radius $a = 15$~mm, outer radius $b = 25$~mm and $t$ as perturbation-depth parameter. Instances of different $t$ values are shown by the dashed/dotted lines (solid line for the unperturbed annulus).  (b) A layout example for one of the manufactured and measured undercoupled feed-driven resonators.}
\label{fig:layout}
\vspace{-\baselineskip}
\end{figure}

Equipped with this analysis of the Helmholtz eigenfunctions for the annulus, we can calculate the matrix entries $A_{ij}$.  Parameterizing the integral in \eqref{matrixentires} in terms of the polar coordinate 
$\theta$, we have $ \rho = -h(\theta)$ and $dS= a d\theta$.
On the inner boundary $\{r=a\}$, 
the tangential derivatives of the separable eigenfunctions with respect to the induced orientation
are
$\partial_{\tau}u_{m,n}^c =\frac{m}{a} R_{m,n}(a) \sin (m \theta)$ and $\partial_{\tau}u_{m,n}^s = -\frac{m}{a} R_{m,n}(a) \cos (m \theta).$
Hence, denoting $u_1 = u_{m,n}^c$, $u_2 = u_{m,n}^s$, 
equation \eqref{matrixentires} in the case of a two-dimensionial eigenspace 
yields
\begin{equation}\label{A12}
    A_{12}=A_{21}=\frac{a}{2}R_{m,n}^2(a) \big(\mu_{m,n} + \frac{m^2}{a^2} \big) \cdot h_{s,2m},
\end{equation}
where we used a double-angle formula to simplify the integral to the Fourier coefficient $h_{s,2m}= \int_{0}^{2 \pi} h(\theta) \cdot \sin(2m \theta) \ d \theta$.
Similarly, the remaining matrix coefficients can be expressed in terms of the Fourier coefficients  $h_0 =  \int_{0}^{2 \pi} h(\theta) d \theta$ and $h_{c, 2m} =  \int_{0}^{2 \pi} h(\theta) \cos(2m \theta) d \theta$.
We find that \begin{equation}\label{A11}A_{11} = \frac{a}{2} R_{m.n}^2(a) \Big[ \big(\mu_{m,n}-\frac{m^2}{a^2}\big) \cdot h_0 + \big(\mu_{m,n}+\frac{m^2}{a^2}\big)\cdot h_{c, 2m} \Big],\end{equation}
\begin{equation}\label{A22} A_{22} = \frac{a}{2} R_{m.n}^2(a) \Big[ \big(\mu_{m,n}-\frac{m^2}{a^2}\big) \cdot h_0 - \big(\mu_{m,n}+\frac{m^2}{a^2}\big) \cdot h_{c, 2m} \Big].\end{equation}
Note the different sign for the second term. In formulas \eqref{A12}--\eqref{A22}, we assume that the functions $R_{m,n}(r)$ of \eqref{radialfunction} are appropriately rescaled so that the corresponding eigenfunctions $u_{m,n}$ are $L^2(\Omega_0)$-normalized. Explicitly, we demand \begin{equation}\label{normalized}\int_a^b rR^2_{m,n}(r) \, dr = 1 / \pi.\end{equation}

Thus, we can control the first order change of several eigenvalues for perturbations of the annulur planar resonator, by prescribing certain Fourier coefficients of the perturbation profile $h(\theta)$.

Indeed, recall that the derivatives of the two eigenvalue branches splitting off from $\mu_{m,n}$ are the eigenvalues of the matrix $(A_{ij})_{1 \leq i,j \leq 2}$. For example, choosing the perturbation $h(\theta)$ such that $h_0 =0, h_{s,2m}=0$ and $h_{c,2m}\neq 0$, the above matrix becomes diagonal of the form $\text{diag}(C, -C)$, and we obtain $\frac{d}{dt}\big|_{t=0} \ \mu_{m,n}^c = -\frac{d}{dt}\big|_{t=0} \ \mu_{m,n}^s$. That is, the
eigenvalue $\mu_{m,n}$ 
splits symmetrically to first order. This is a key parameter for resonator design in practice. 
Picking the perturbation profile $h(\theta) = \cos(2 m\theta)$ then yields
\begin{equation}\label{firstderivative}
\frac{d}{dt}\Big|_{t=0} \ \mu_{m,n}^c =  \frac{\pi a R_{m,n}^2(a)}{2} \big(\mu_{m,n} + \frac{m^2}{a^2} \big).
\end{equation}
Taking $m=1, n=1$, we can split the smallest non-zero eigenvalue symmetrically at first order.


\section{Applied Examples}\label{sec: examples}

An annular stripline resonator perturbed by the series of shapes described by \eqref{perturbations} with $h(\theta)=\cos(2 \theta)$ was simulated using the independent FEM eigen solver in Ansys HFSS. The resulting set of perturbed resonators were manufactured and their frequencies measured. An FR408HR substrate with $\varepsilon_r\approx3.69$ and loss tangent $\tan\delta\approx0.0092$ was used, with 1.5~mm heights (top and bottom), feed-lines width of 3.3~mm, feed-head flare angles of 20~degrees and a feed gap of 0.2~mm that provides sufficient undercoupling from the resonator for accurate measurements. We fixed the aspect ratio to be $b/a=5/3$, set $a=15$~mm and varied $t$ from 0 to 6~mm (see Figure~\ref{fig:layout}). Good agreement is observed between measurements, simulations, and linear extrapolation from our theory. \\

\begin{table}[htbp]
\centering
\begin{tabular}{c|ccccc}
\hline
$m$       & 0 & 1 & 2 & 3 & 4  \\
\hline
$k_{m,1}$ & 317.204 &  50.483 & 100.638 & 150.158 & 198.786  \\
$k_{m,2}$ & 629.891 & 321.644 & 334.683 & 355.573 & 383.294  \\
$k_{m,3}$ & 943.533 & 632.042 & 638.459 & 649.033 & 663.599  \\
\hline
\end{tabular}
\par\vspace{6pt}
\caption{The first few positive zeros $k_{m,n}$ of equation \eqref{equationforks} for the Neumann annulus with $a=15$~mm and $a/b = 3/5$.}
\label{tab:annulus-wavenumbers/roots-a3cm}
\end{table}

Table \ref{tab:annulus-wavenumbers/roots-a3cm} shows the values of the zeros $k_{m,n}$ of defining equation \eqref{equationforks} for $0{\leq}m {\leq} 4, \ 1 {\leq} n {\leq} 3$. The first few zeros are
\begin{equation*}
    k_{1,1}{<}k_{2,1}{<}k_{3,1}{<}k_{4,1}{<}k_{5,1}{<}k_{6,1} {<}k_{0,1}{<}k_{1,2}{<}k_{2,2}{<}\cdots,
\end{equation*}
in that order.
\begin{figure}[!t]
\centering
\includegraphics[width=75mm]{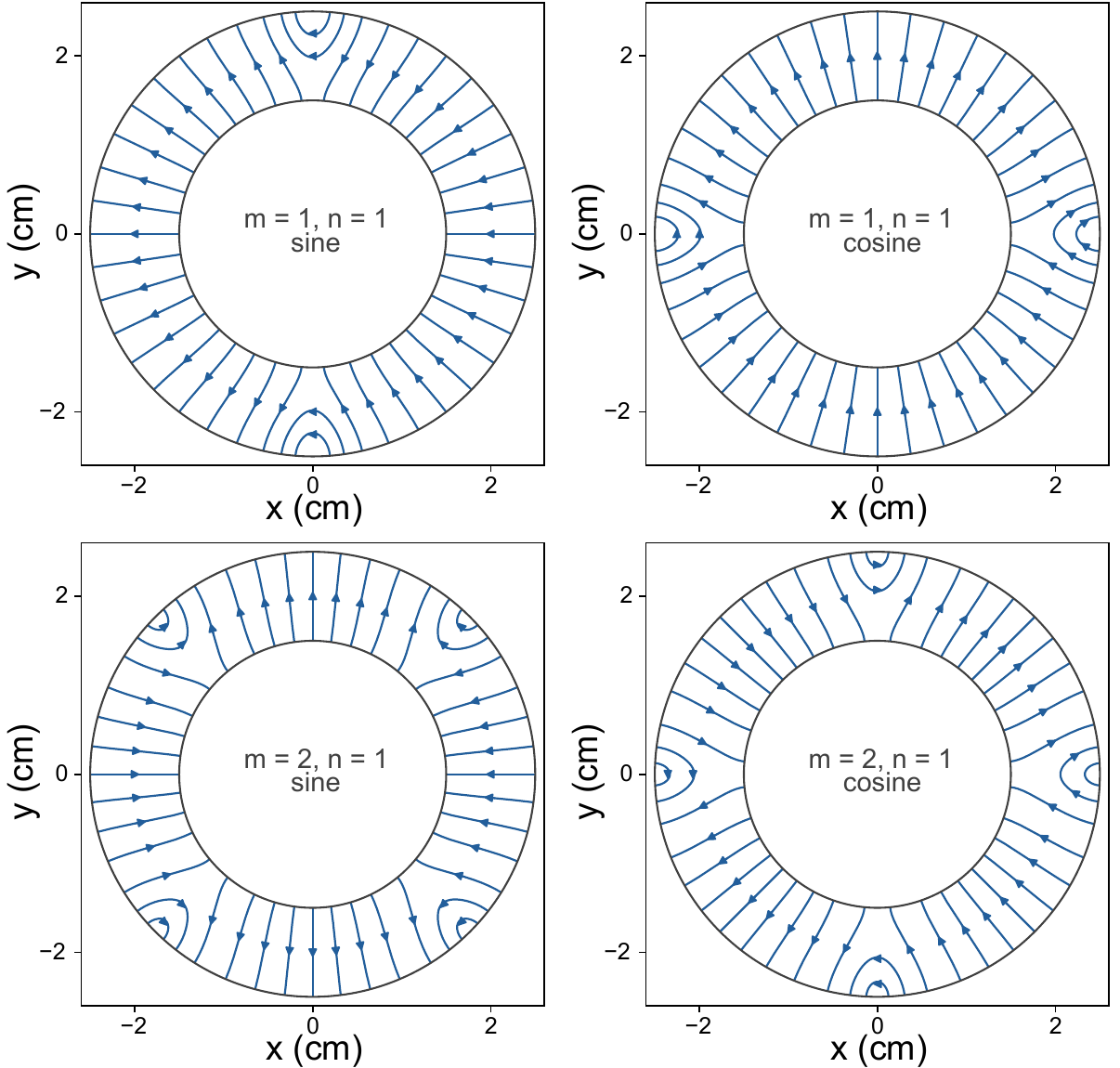}
\caption{H-field streamline plots for the first four TM$^z$ eigenmodes of unperturbed annulus, shown here for $a=15$~mm and $a/b = 3/5$. }
\label{fig:Hfield}
\vspace{-\baselineskip}
\end{figure}
Using $f=k c/(2\pi)$ and  $\varepsilon_{r}\approx3.69$, our theory predicts the first few resonant frequencies of the unperturbed annular resonator to be 
$
f_{1,1} = 1.254\,\mathrm{GHz},  f_{2,1} = 2.500\,\mathrm{GHz}, f_{3,1} = 3.730 \,\mathrm{GHz},
$
all of them having multiplicity two. 
The magnetic field $\mathbf{H}$ of an eigenmode solution, which can be obtained from $E_z$ by \eqref{Hcomponents}, is uniform in $z$. Figure \ref{fig:Hfield} shows the streamline plot of the magnetic field for the first four eigenmodes, generated using Wolfram's Mathematica. The expected azimuthal and radial orders
of the first two Helmholtz eigenvalues $k_{1,1}^2<k_{2,1}^2$ (each having multiplicity two) are observed.

\begin{figure}[t]
\centering
\includegraphics[width=88mm]{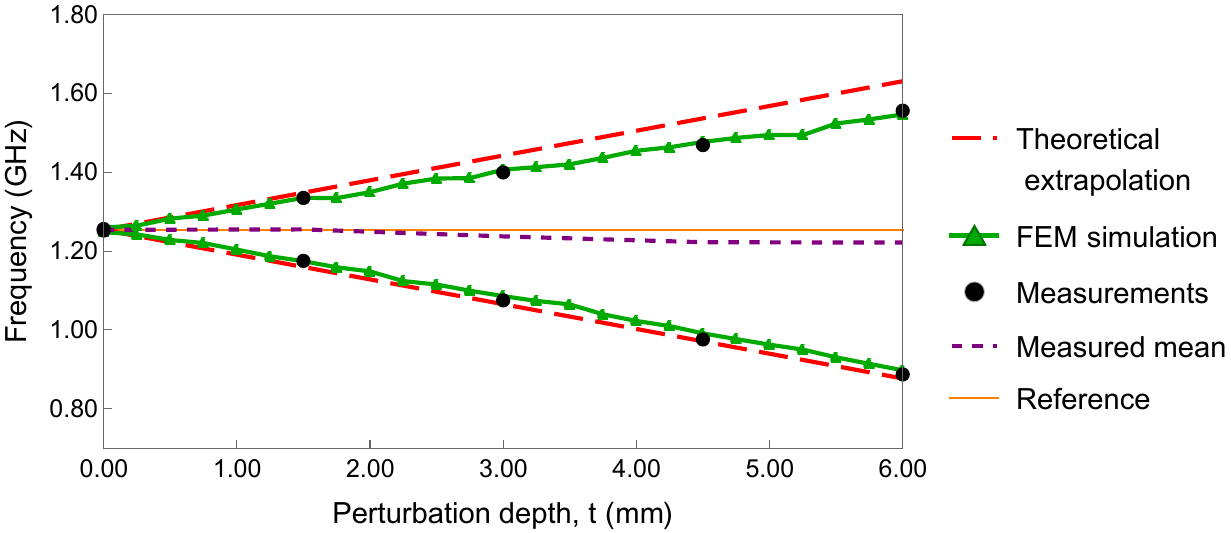}
\caption{Frequency splitting as a function of perturbation depth, $t$. Measurements are compared wtih FEM simulations and theoretical extrapolation. }
\label{fig:perturbedfrequencyplot}
\vspace{-\baselineskip}
\end{figure}

Our theory predicts the rate of change of the first two resonant frequencies $f_{1,1}^s, f_{1,1}^c$ via equation \eqref{firstderivative} and the chain rule 
\begin{equation}\label{chainrule}\frac{d}{dt}\Big|_{t=0}f^c_{1,1}(t) = \frac{c}{4 \pi \sqrt{\mu_{1,1}}} \frac{d}{dt} \Big|_{t=0}\mu_{1,1}^c(t).\end{equation}
In order to produce correct numerical predictions from this theory, we need to normalize the radial function $R_{1,1}(r)$ given by \eqref{radialfunction}  so as to satisfy \eqref{normalized}. 
We do so by noting that
$$\int_{a}^b rR_{1,1}^2(r) dr = \frac{1}{2} \Big[\big( r^2 - 1/k_{1,1}^2 \big)R_{1,1}^2(r) \Big]_a^b,$$
because $R_{1,1}(r)$ solves the first-order Bessel equation and $R_{1,1}'(a)=R_{1,1}'(b)=0$. Using numerical values of the Bessel functions, one finds that the function $R_{1,1}(r)$ given by \eqref{radialfunction} is to be rescaled by a factor of
$46.8499893$.

The two frequencies $f_{1,1}^s, f_{1,1}^c$ are observed in practice as frequency shifts above and below the unperturbed frequency value. For the specific radii and substrate chosen, equations \eqref{firstderivative} and \eqref{chainrule} predict the rate of change of the first two resonant frequencies to be $\pm 62.9$~MHz/mm. Figure~\ref{fig:perturbedfrequencyplot} and Table~\ref{tab:data} show the splitting of the frequency of the first eigenmode, TM$^z_{11}$, as a function of the perturbation depth $t$. The splitting behavior is due to the perturbation shapes $t\cos(2\theta)$ suggested by our theory. The unperturbed frequency at the instant $t=0$ is $f_0=1.254$~GHz. 
The values extrapolated from the first-order shape derivative in \eqref{firstderivative} and \eqref{chainrule} are compared with the independent HFSS simulation results and measurements.
Good agreement is observed, with a maximum deviation of about 5\% between measurement and theoretical extrapolation at $t=6$~mm.  It is interesting to observe that the local first-order prediction by the theory actually persists to a relatively wide tuning range, from $0.89$~GHz to $1.56$~GHz, at $t=6$~mm, which corresponds to $(f^c_{1,1}-f^s_{1,1})/f_0=54$\%. This increases the usefulness of the suggested method for practical design.


\begin{table}[htbp]
\centering
\begin{tabular}{c|cccc}
\hline
$t$ & Theory extrapolation & FEM Simulations  & Measurements \\ 
(mm)& $f^s_{1,1},f^c_{1,1}$~(GHz)& $f^s_{1,1},f^c_{1,1}$~(GHz)& $f^s_{1,1},f^c_{1,1}$~(GHz) \\
\hline
0.0 & 1.25, 1.25 & 1.25, 1.25 & 1.25, 1.26 \\
1.5 & 1.16,	1.35 &	1.16, 1.34 & 1.18, 1.33 \\
3.0 & 1.07,	1.44 &	1.07, 1.40 & 1.07, 1.40 \\
4.5 & 0.97, 1.54 &	1.00, 1.46 & 0.98, 1.47 \\
6.0 &  0.88, 1.63 &	0.90, 1.53 & 0.89, 1.56 \\
\hline
\end{tabular}
\par\vspace{6pt}
\caption{Data for split frequencies, $f^s_{1,1}$ and $f^c_{1,1}$, as a function of perturbation depth, $t$. See corresponding plot in Figure~\ref{fig:perturbedfrequencyplot}. 
}
\label{tab:data}
\end{table}


\section{Conclusion}

A mathematical framework is outlined for the spectral manipulation of eigenmodes in a planar resonator through prescribed geometric perturbations. The framework is used to suggest a solution for the specific problem of splitting a degenerate eigenmode symmetrically while keeping the center frequency stationary.
Simulations and measurements show good agreement with the theory and demonstrate that the local first-order prediction made by the theory persists to a relatively wide tuning range. This increases its value for practical design.   



\section*{Acknowledgment}

The authors thank Evan Gawlik, Santa Clara University, and Rafe Mazzeo, Stanford University, for valuable discussions related to this work. This work is supported by the National Science Foundation, USA (ID: 2501494).


\bibliographystyle{IEEEtran}

\bibliography{IEEEabrv,EigenmodeSplittingBib}

@ARTICLE{Hong_triangle,
  author={Jia Sheng Hong and Shuzhou Li},
  journal={IEEE Transactions on Microwave Theory and Techniques}, 
  title={Theory and experiment of dual-mode microstrip triangular patch resonators and filters}, 
  year={2004},
  volume={52},
  number={4},
  pages={1237-1243},
  doi={10.1109/TMTT.2004.825653}
  }

@ARTICLE{Gorur,
  author={Gorur, A.},
  journal={IEEE Transactions on Microwave Theory and Techniques}, 
  title={Description of coupling between degenerate modes of a dual-mode microstrip loop resonator using a novel perturbation arrangement and its dual-mode bandpass filter applications}, 
  year={2004},
  volume={52},
  number={2},
  pages={671-677},
  doi={10.1109/TMTT.2003.822033}}

@ARTICLE{Lin,
  author={Lin, Tsu Wei and Kuo, Jen Tsai and Chung, Shyh Jong},
  journal={IEEE Transactions on Microwave Theory and Techniques}, 
  title={Dual-Mode Ring Resonator Bandpass Filter With Asymmetric Inductive Coupling and Its Miniaturization}, 
  year={2012},
  volume={60},
  number={9},
  pages={2808-2814},
  doi={10.1109/TMTT.2012.2205936}}

@article{Naji_stability,
  author    = {Naji, Adham and Soliman, Mina H.},
  title     = {Center Frequency Stabilization in Planar Dual-Mode Resonators during Mode-Splitting Control},
  journal   = {Scientific Reports},
  year      = {2017},
  volume    = {7},
  number    = {1},
  pages     = {43855},
  month     = {mar},
  doi       = {10.1038/srep43855}
}

@ARTICLE{Naji_geo,
  author={Naji, Adham and Warr, Paul and Beach, Mark and Morris, Kevin},
  journal={IEEE Transactions on Microwave Theory and Techniques}, 
  title={A Fundamental Limit on the Performance of Geometrically-Tuned Planar Resonators}, 
  year={2011},
  volume={59},
  number={6},
  pages={1491-1499},
  doi={10.1109/TMTT.2011.2118227}}

@book{Balanis,
  title={Antenna Theory: Analysis and Design},
  author={Balanis, Constantine A.},
  edition={4th},
  year={2016},
  publisher={John Wiley \& Sons},
  address={Hoboken, NJ},
  isbn={978-1-118-64206-1}
}

@book{Collin,
isbn = {0879422378},
publisher = {IEEE Press},
title = {Field theory of guided waves},
year = {1991},
author = {Collin, Robert E.},
address = {New York},
booktitle = {Field theory of guided waves},
edition = {2nd ed.},
}

@book{Delfour,
  author    = {Delfour, Michel C. and Zol{\'e}sio, Jean Paul},
  title     = {Shapes and Geometries: Metrics, Analysis, Differential Calculus, and Optimization},
  edition   = {2nd},
  publisher = {Society for Industrial and Applied Mathematics (SIAM)},
  year      = {2011},
  series    = {Advances in Design and Control},
  volume    = {22},
  address   = {Philadelphia, PA},
  isbn      = {978-0-89871-936-9},
}

@book{Hong,
  title     = {Microstrip Filters for RF/Microwave Applications},
  author    = {Hong, Jia Sheng and Lancaster, Michael J.},
  edition   = {2nd},
  year      = {2011},
  publisher = {John Wiley \& Sons},
  address   = {Hoboken, New Jersey},
  isbn      = {978-0-470-40877-3}
}

@book{Cameron,
  author    = {Richard J. Cameron and Chandra M. Kudsia and Raafat R. Mansour},
  title     = {Microwave Filters for Communication Systems: Fundamentals, Design, and Applications},
  publisher = {John Wiley \& Sons},
  year      = {2018},
  edition   = {2nd},
  address   = {Hoboken, NJ},
  isbn      = {978-1-118-27434-7}
}

@book{Slater,
  author    = {Slater, John C.},
  title     = {Microwave Electronics},
  publisher = {D. Van Nostrand Company, Inc.},
  address   = {Princeton, NJ},
  year      = {1950},
  pages     = {xiv + 406},
  series    = {The Bell Telephone Laboratories Series}
}

@article{Lamberti,
author = {Lamberti, Pier Domenico and Zaccaron, Michele},
title = {Shape sensitivity analysis for electromagnetic cavities},
journal = {Mathematical Methods in the Applied Sciences},
volume = {44},
number = {13},
pages = {10477-10500},
year = {2021}
}

@book{Henry,
  title     = {Perturbation of the Boundary in Boundary-Value Problems of Partial Differential Equations},
  author    = {Henry, Dan},
  series    = {London Mathematical Society Lecture Note Series},
  volume    = {318},
  year      = {2005},
  publisher = {Cambridge University Press},
  address   = {Cambridge, UK},
  isbn      = {9780511546730}
}

\end{document}